\documentclass[unsortedaddress,superscriptaddress,prl,10pt,aps,twocolumn,showpacs]{revtex4-2}
\usepackage[pdftex]{graphicx}
\usepackage{graphicx}
\usepackage[latin1]{inputenc}
\usepackage{amsmath}
\usepackage{ulem}
\usepackage{bbold}
\usepackage{hyperref}
\usepackage{url}
\usepackage{wasysym}
\usepackage[T1]{fontenc}
\usepackage{color,bm, braket}
\usepackage[]{siunitx}

\hypersetup{
    colorlinks,
    citecolor=black,
    filecolor=black,
    linkcolor=black,
    urlcolor=black
}

\usepackage{xcolor}
\usepackage{upgreek}
\newcommand {\mo}{MoS$_2$}

\begin{document}
\DeclareGraphicsExtensions{.pdf}

\title{Tip-induced creation and Jahn-Teller distortions of\\ sulfur vacancies in single-layer \mo{}
}

\author{Daniel Jansen} 
\affiliation{II. Physikalisches Institut, Universit\"{a}t zu K\"{o}ln, Z\"{u}lpicher Stra\ss e 77, 50937 K\"{o}ln, Germany}
\author{Tfyeche Tounsi}
\affiliation{II. Physikalisches Institut, Universit\"{a}t zu K\"{o}ln, Z\"{u}lpicher Stra\ss e 77, 50937 K\"{o}ln, Germany}
\author{Jeison Fischer}
\affiliation{II. Physikalisches Institut, Universit\"{a}t zu K\"{o}ln, Z\"{u}lpicher Stra\ss e 77, 50937 K\"{o}ln, Germany}
\author{Arkady V. Krasheninnikov}
\affiliation{Institute of Ion Beam Physics and Materials Research, Helmholtz-Zentrum Dresden-Rossendorf, 01328 Dresden, Germany}
\author{Thomas Michely}
\affiliation{II. Physikalisches Institut, Universit\"{a}t zu K\"{o}ln, Z\"{u}lpicher Stra\ss e 77, 50937 K\"{o}ln, Germany}
\author{Hannu-Pekka Komsa}
\affiliation{Faculty of Information Technology and Electrical Engineering, University of Oulu, FI-90014 Oulu, Finland}
\author{Wouter Jolie}
\affiliation{II. Physikalisches Institut, Universit\"{a}t zu K\"{o}ln, Z\"{u}lpicher Stra\ss e 77, 50937 K\"{o}ln, Germany}

\begin{abstract}

We present an atomically precise technique to create sulfur vacancies and control their atomic configurations in single-layer \mo{}. It involves adsorbed Fe atoms and the tip of a scanning tunneling microscope, which enables single sulfur removal from the top sulfur layer at the initial position of Fe. Using scanning tunneling spectroscopy, we show that the STM tip can also induce two Jahn-Teller distorted states with reduced orbital symmetry in the sulfur vacancies. Density functional theory calculations rationalize our experimental results. Additionally, we provide evidence for molecule-like hybrid orbitals in artificially created sulfur vacancy dimers, which illustrates the potential of our technique for the development of extended defect lattices and tailored electronic band structures.

\end{abstract}

\maketitle

Point defects in semiconducting two-dimensional (2D) transition metal dichalcogenides (TMDs) yield atomically precise quantum states of matter. Defects can induce isolated in-gap states \cite{schuler2019large, zhu2021dimensionality, trainer2022visualization}, which are envisioned to have potential as qubits \cite{zhang2020material,li2022carbon} or single-photon emitters \cite{gupta2018two,zhang2020material}. Having access to such isolated levels enables detection of physical properties related to nano-magnetism \cite{zhang2019engineering,cochrane2021spin,trishin2023electronic}, spin-orbit coupling \cite{schuler2019substitutional}, or chemical bond formation \cite{vancso2016intrinsic, schuler2019substitutional, zhao2023electrical}.

One way to create an isolated defect is via atom removal. Vacancies can be introduced in TMDs through impacts of energetic electrons in transmission electron microscopy \cite{zhou2013intrinsic,komsa2013point,hong2015exploring}, ion irradiation \cite{mitterreiter2020atomistic, valerius2020reversible, mitterreiter2021role, ozden2023engineering},  as well as via annealing in ultrahigh vacuum (UHV) \cite{schuler2019large}. While the concentration of vacancies can be controlled to some extent via ion/electron irradiation fluence and particle energy, annealing time and sample temperature, these techniques tend to create other types of defects as well, often with unknown atomic structure. The elevated temperatures can also lead to diffusion and agglomeration of vacancies \cite{fang2023atomically}. Hence, a method able to locally create and probe a single type of vacancy in 2D TMDs remains elusive. It would allow unambiguous identification of theoretical predictions, such as a Jahn-Teller distortion when an extra electron occupies one of the in-gap states~\cite{komsa2015native,gupta2018two,tan2020stability}.

\begin{figure}[b]
	\centering
		\includegraphics[width=0.495\textwidth]{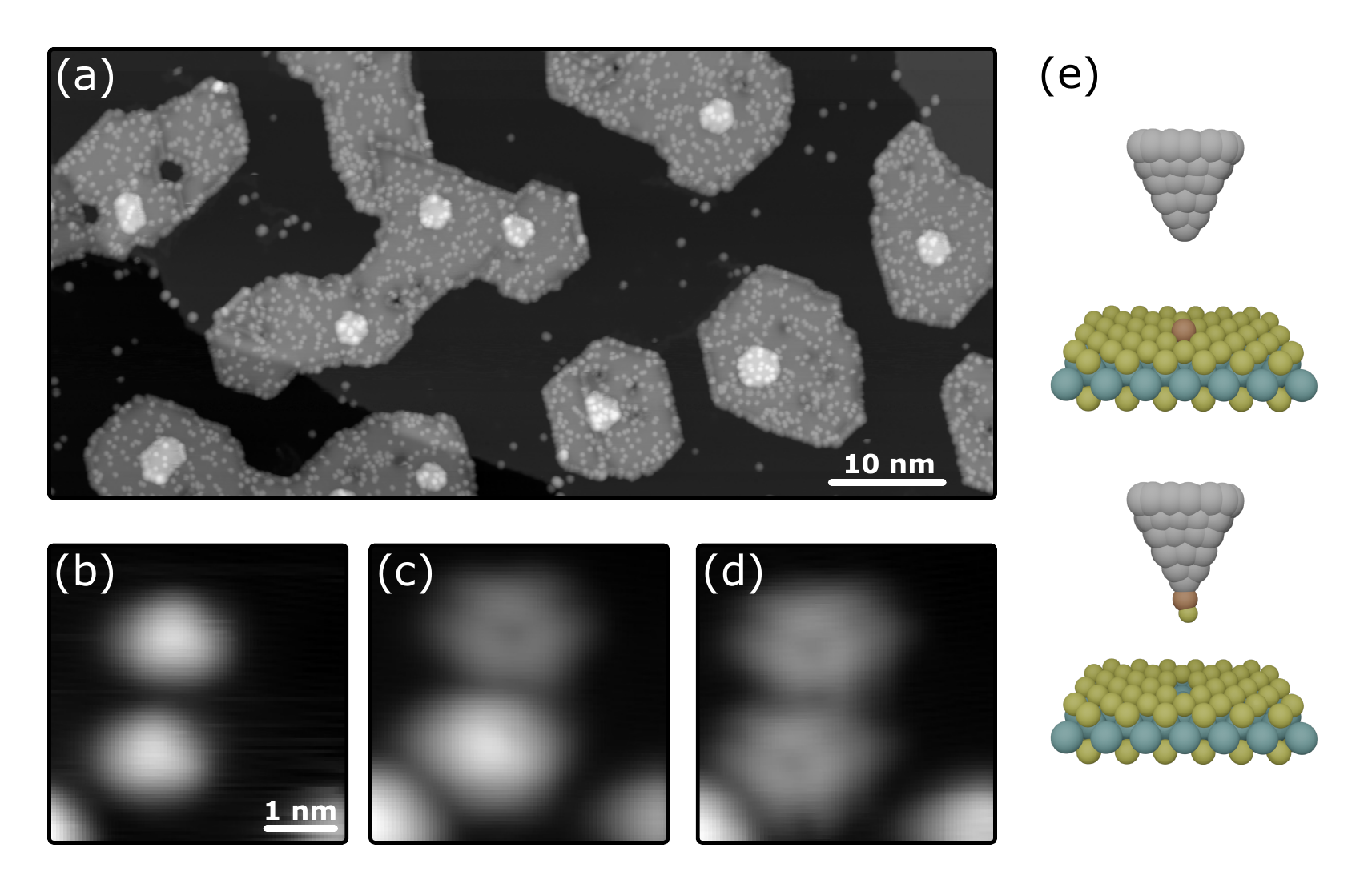}
	\caption{Local creation of sulfur vacancies. (a) STM image of single-layer and bilayer \mo{} islands on Gr/Ir(111) covered with Fe adatoms ($V_{\text{bias}}=$ 1 V, $I_{\text{set}}=$ 100 pA, image sizes: 200$\times$90 nm$^{2}$). (b)-(d) Step-by-step creation of sulfur vacancies. (b) Two adjacent Fe adatoms on \mo{} ($V_{\text{bias}}=$ 1 V, $I_{\text{set}}=$ 100 pA, image size: 4$\times$4 nm$^{2}$). (c) Upper Fe adatom removed and sulfur vacancy created at its location ($V_{\text{bias}}=$ 500 mV, $I_{\text{set}}=$ 100 pA). (d)  Second Fe adatom removed and  vacancy created ($V_{\text{bias}}=$ 500 mV, $I_{\text{set}}=$ 100 pA). (e) Sketch visualizing how the STM tip creates sulfur vacancies in \mo{}.}
	\label{figure 1}
\end{figure}

\begin{figure*}[t]
	\centering
		\includegraphics[width=0.7\textwidth]{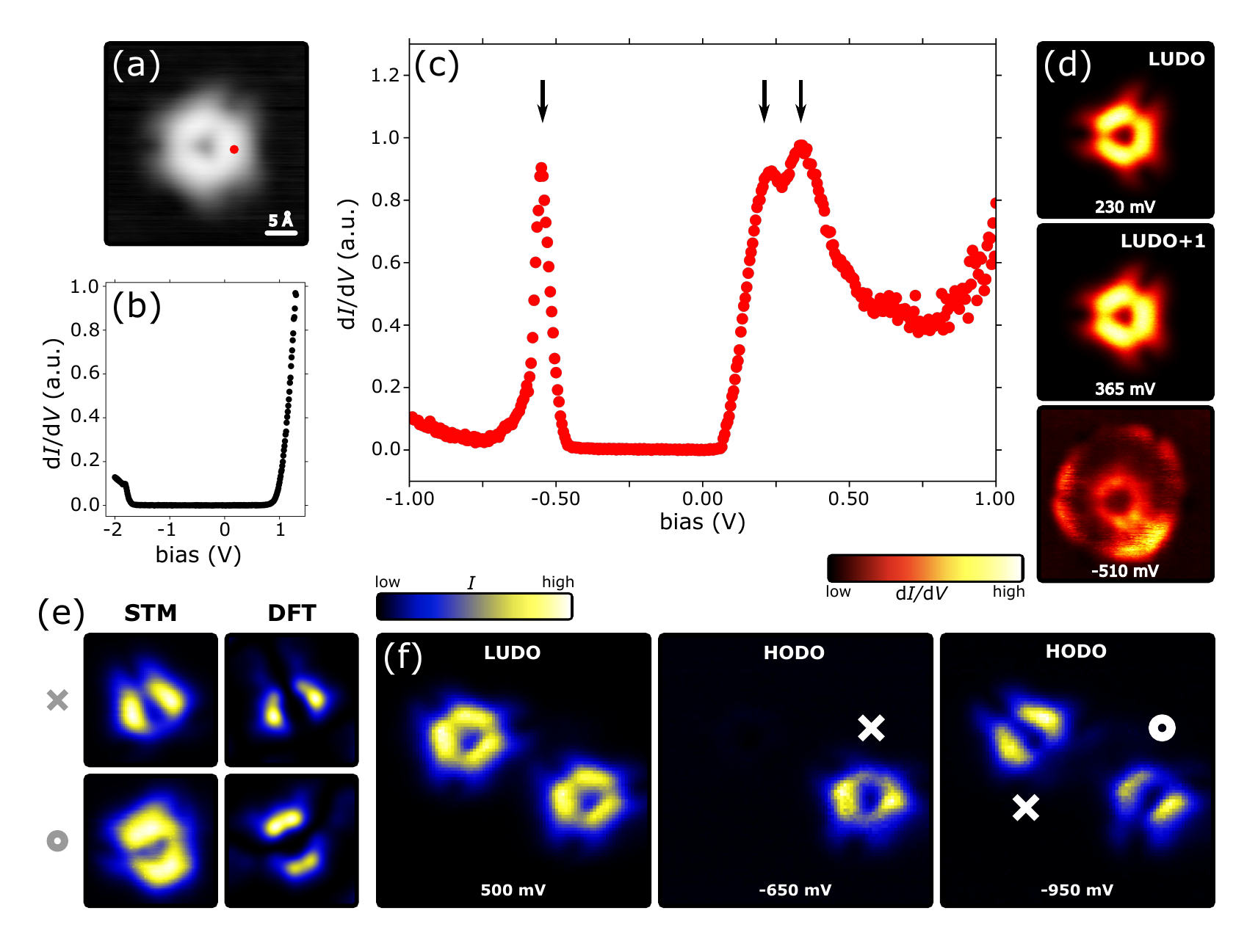}
	\caption{Spectroscopic investigation of sulfur vacancies. (a) STM image of a created sulfur vacancy ($V_{\text{bias}}=$ 1 V, $I_{\text{set}}=$ 100 pA, image size: 3$\times$3 nm$^{2}$ ). The red spot marks the location where the spectrum presented in (c) was measured. (b) d$I$/d$V$ spectrum ($V_{\text{stab}}=$ 1.3 V, $I_{\text{stab}}=$ 100 pA, $V_{\text{mod}}=$ 10 mV) measured on pristine \mo{}. (c) d$I$/d$V$ spectrum ($V_{\text{stab}}=$ 1 V, $I_{\text{stab}}=$ 500 pA, $V_{\text{mod}}=$ 2 mV) measured on the vacancy, revealing three peaks in the band gap of \mo{}. (d) Constant-height d$I$/d$V$ maps (size: 3$\times$3 nm$^{2}$) measured at energies indicated by black arrows in (c). (e) Constant-height current maps of the two distinct symmetry-broken HODO orbitals and corresponding DFT simulations. The two types of orbitals are denoted by crosses ($\times$) and circles ($\ocircle$). (f) Series of current maps showing transitions between the two symmetry-broken states (image sizes: 7$\times$7 nm$^{2}$).}
	\label{figure 2}
\end{figure*}

In this letter, we employ an atomic manipulation approach to create single vacancies in the top sulfur layer of single-layer \mo{}. We use Fe atoms adsorbed on \mo{} as atomic-scale markers for vacancy creation. A detailed characterization of isolated vacancies reveals two Jahn-Teller distorted states when the vacancy is charged. Creating a vacancy dimer shows the formation of hybridized states, while no additional distortion is observed upon charging.

All scanning tunneling microscopy (STM) and spectroscopy (STS) measurements were performed in a UHV low-temperature STM setup with an operating temperature of 6 K. \mo{} grown on Gr on Ir(111) is quasi-freestanding, as evidenced by its large band gap~\cite{murray2019comprehensive}, sharp valence band~\cite{ehlen2018narrow} and narrow photoluminescence peak~\cite{ehlen2018narrow}. The Ir(111) crystal was first cleaned by cycles of 1.5 keV Ar$^{+}$ ion beam exposure and annealed to 1550 K. Subsequently, single crystal Gr was grown using the growth method described in Ref. \cite{coraux2008structure}. \mo{} islands were grown according to Ref.~\cite{hall2018molecular} using an e-beam evaporator for Mo and a Knudsen cell for sulfur supply. The \mo{} islands were grown at room temperature with a Mo flux of 1.14 $\cdot$ 10$^{16}$ $\frac{\text{atoms}}{\text{m}^{2} \cdot \text{s}}$, a deposition time of 600~s and a sulfur pressure of 6 $\cdot$ 10$^{-9}$~mbar. Afterwards the sample was annealed in a sulfur pressure of 3 $\cdot$ 10$^{-9}$~mbar at 1050~K for 300~s, resulting in virtually defect-free single-layer \mo{} islands with small second layer islands on top. Single Fe atom evaporation was performed using an e-beam evaporator at a sample temperature of 8 K. An STM image of the as-grown sample is presented in Fig. \ref{figure 1}(a). It becomes apparent, that Fe on \mo{} is non-mobile and does not cluster at low Fe densities.

Sulfur vacancies can be created with high reproducibility by the following procedure: First, the STM tip is positioned over an Fe adatom and approached by decreasing the bias voltage $V_{\text{bias}}$ and increasing the tunneling current $I_{\text{set}}$ with closed feedback loop. The tip-adatom separation is reduced until a sudden change in the measured tip height is detected. Typical values are $V_{\text{bias}}\approx$ 30 mV and $I_{\text{set}}\approx$ 1 nA. After this event, the tip height is increased to scanning parameters (typically $V_{\text{bias}}=$ 1 V and I$_{\text{set}}=$ 100 pA). In the large majority of attempts, this procedure creates a sulfur vacancy. A series of STM images showing the subsequent creation of two neighboring sulfur vacancies is presented in Fig. \ref{figure 1}(b)-(d). 

A model interpreting the observations is shown in Fig. \ref{figure 1}(e). The pick-up event involves the adsorbed Fe atom together with a single sulfur atom, leaving a sulfur vacancy in \mo{} behind. According to our DFT calculations, the Fe-sulfur bond alone is not sufficient to create a vacancy. Additional effects, such as the electric field between tip and surface \cite{moradi2024strain}, as well as vibrations induced by tunneling electrons can facilitate pick-up, though a full picture of the mechanism lies beyond the scope of the present study. While the chemical bond between tip and Fe strongly depends on apex geometry and chemical composition, the Fe-sulfur bond is well defined for every pick-up event. This leads to high reproducibility of the vacancy creation process, as long as the bond between Fe and STM tip is strong enough to initiate the pick-up event. This is in contrast to direct sulfur pick-up events without Fe adatoms reported earlier for bulk \mo{} \cite{hosoki1992surface}, which we found not to be applicable to single-layer \mo{} islands.

The electronic characterization of an isolated sulfur vacancy created using our atomic manipulation approach is presented in Fig. \ref{figure 2}. An STM image of a single vacancy is shown in Fig. \ref{figure 2}(a). Its appearance depends on the applied bias voltage, which is discussed in Fig. S1. An STS point spectrum acquired on pristine \mo{} [Fig. \ref{figure 2}(b)] displays its band gap \cite{murray2019comprehensive}. Point spectra taken on the defect lobes reveal the existence of three peaks in the band gap of \mo{}, as shown in Fig. \ref{figure 2}(c). Corresponding d$I$/d$V$ maps measured at the respective peak energies above the Fermi level and presented in Fig. \ref{figure 2}(d) show two nearly-equivalent orbitals with C$_{\text{3v}}$ symmetry, the lowest unoccupied defect orbitals LUDO and LUDO+1. In contrast, the d$I$/d$V$ map measured close to the peak energy below the Fermi level reveals a modified orbital shape, surrounded by a pronounced ring. 

The origin of the peak in the occupied states and the ring in the d$I$/d$V$ map is a charging event induced by the STM tip \cite{pradhan2005atomic}. Applying a bias voltage between tip and sample %creates an electric field, locally gating the \mo{} \cite{mcellistrem1993electrostatic, zhang2012band}. This 
implies an electric field between them and thus causes the in-gap states of the sulfur vacancy to shift in energy. For large enough negative bias voltages, the LUDO level is pulled below the Fermi energy and occupied by an electron, thus becoming the highest occupied defect orbital (HODO) of the charged vacancy. The peak in the point spectrum of Fig. \ref{figure 2}(c) and the ring in the d$I$/d$V$ map in Fig. \ref{figure 2}(d) at -510 mV thus reflect a sudden change in the occupied defect states due to charging and subsequent distribution of that charge. Similar results have previously been obtained for sulfur vacancies in single-layer WS$_2$ \cite{schuler2019large}.

\begin{figure}[tb]
	\centering
		\includegraphics[width=0.495\textwidth]{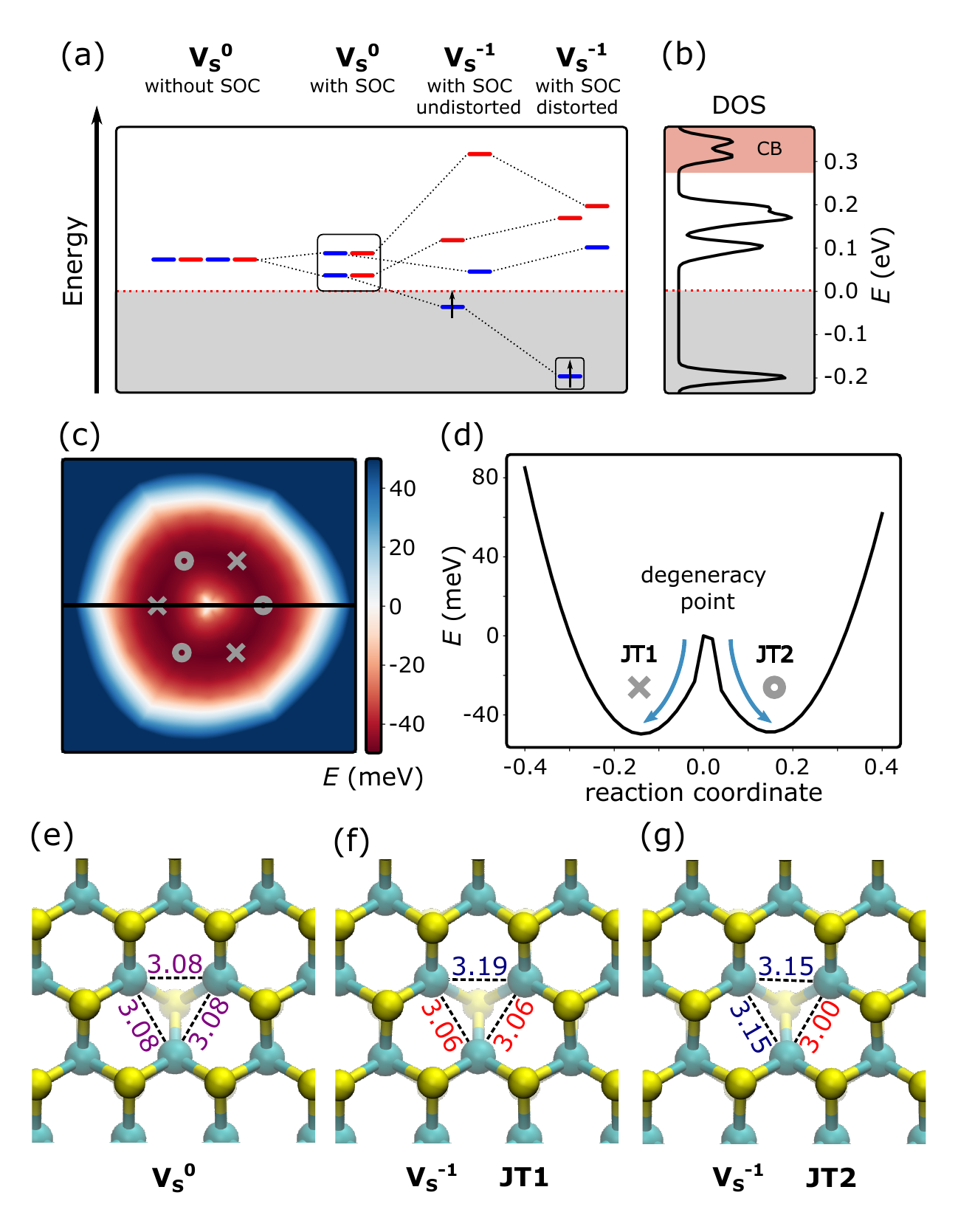}
	\caption{Jahn-Teller distortions in negatively charged sulfur vacancies. (a) Evolution of sulfur vacancy levels (blue: spin-up, red: spin-down) upon introduction of spin-orbit coupling, charge and lattice distortions. The gray shaded area indicates state occupation and the red dashed line the Fermi level. Black solid boxes emphasize which states are accessible in experiment. (b) DFT calculated DOS of the distorted negatively charged sulfur vacancy. (c) Potential energy surface of the negatively charged sulfur vacancy. Upon charging, the vacancy can relax into two different JT distorted states indicated by crosses ($\times$) and circles ($\ocircle$), respectively. (d) Slice through the potential energy surface along black line displayed in (c). (e)-(g) Structure models of the \mo{} lattice around the sulfur vacancy for the neutral (e) and charged case in JT1 (f) and JT2 (g) states. The numbers on the dashed lines give the Mo atom separations ($\text{\AA}$).}
	\label{figure 3}
\end{figure}

The absence of additional states in the \mo{} band gap enables us to access the HODO orbital at high bias voltages at which the vacancy is stably charged, by measuring the tunneling current in constant-height mode. This signal integrates over all states between the Fermi energy and the bias voltage, which in our system is dominated by the HODO only, see Fig. S2 in the Supplemental Material for detailed reasoning why constant-height current maps are used to image the HODO. Constant-height maps recording the tunneling current of two representative vacancies are presented in Fig. \ref{figure 2}(e). The high-resolution maps reveal two different types of HODOs, denoted as ($\times$) and ($\ocircle$). Both show a clear symmetry reduction from C$_{\text{3v}}$ to C$_{\text{S}}$. A statistical analysis of 16 vacancies reveals that 10 of the HODOs have the appearance ($\times$), while the remaining 6 display the ($\ocircle$) orbital shape. These orbitals can transform from one to another when neighboring vacancies are charged. This is demonstrated in Fig. \ref{figure 2}(f) for two neighboring vacancies. Their LUDOs are measured at 500 mV. At -650 mV, only one of the vacancies is charged, due to differences in the in-gap state energies of the neutral vacancies, displayed in Fig. S3. Hence, only one HODO is observed, in state ($\times$). Increasing the bias voltage eventually charges the second vacancy. While the latter is in state ($\times$), the other transformed into state ($\ocircle$).

The behavior of the vacancies upon charging can be understood with the help of density functional theory (DFT). For details on the computational methods we refer to the Supplemental Material. Negatively charged sulfur vacancies in \mo{} have been predicted to undergo a JT effect, which distorts the host lattice and spontaneously breaks the (orbital) symmetry from C$_{\text{3v}}$ to C$_{\text{S}}$ \cite{komsa2015native, gupta2018two, tan2020stability}. To understand the JT effect in \mo{} sulfur vacancies, we simulate the energy levels of a \mo{} vacancy, see Fig. \ref{figure 3}(a). The three dangling bonds from Mo atoms ($\phi_a$, $\phi_b$, and $\phi_c$) in $C_{3v}$ symmetry lead to the following symmetry-adapted linear combinations (SALCs):
$\phi_1 = \phi_a + \phi_b + \phi_c$,
$\phi_2 = 2\phi_a - \phi_b - \phi_c$, and
$\phi_3 = \phi_b - \phi_c$, where
$\phi_1$ belongs to $a_1$ and $\phi_2$ and $\phi_3$ to $e$. The lowest $a_1$ state is occupied by two electrons and lies close to the VBM [not shown in Fig. \ref{figure 3}(a)]. The $a_1$ state is thus too far away from the Fermi level to be depopulated by the electric field of the tip. The two degenerate $e$ states are unoccupied and fall in the band gap of \mo{} \cite{pandey2016defect}, see Fig. \ref{figure 3}(a) (V$_{\text{S}}^{0}$ without SOC). The SALCs can be clearly seen in the charge density plots presented in Fig. S4. Adding SOC partially lifts the degeneracy of the $e$ states, leading to two doubly-degenerate levels [Fig. \ref{figure 3}(a) (V$_{\text{S}}^{0}$ with SOC)] \cite{khan2017electronic}. These levels correspond to the LUDO and LUDO+1 orbitals found in experiment. Adding an electron to the lowest level lifts all remaining degeneracies, leaving one uncompensated spin below the Fermi energy [Fig. \ref{figure 3}(a) (V$_{\text{S}}^{-1}$ with SOC undistorted)]. Relaxing the structure spontaneously distorts the atomic configuration around the vacancy, leading to the energy level configuration of the JT effect in negatively charged \mo{} sulfur vacancies, presented in Fig. \ref{figure 3}(a) (V$_{\text{S}}^{-1}$ with SOC distorted). The corresponding density of states (DOS) is presented in Fig. \ref{figure 3}(b). Note that only the HODO is experimentally accessible [solid black box in the level scheme in Fig. \ref{figure 3}(a)], since we need to apply a negative bias voltage to charge the vacancy.

\begin{figure}[tb]
	\centering
		\includegraphics[width=0.495\textwidth]{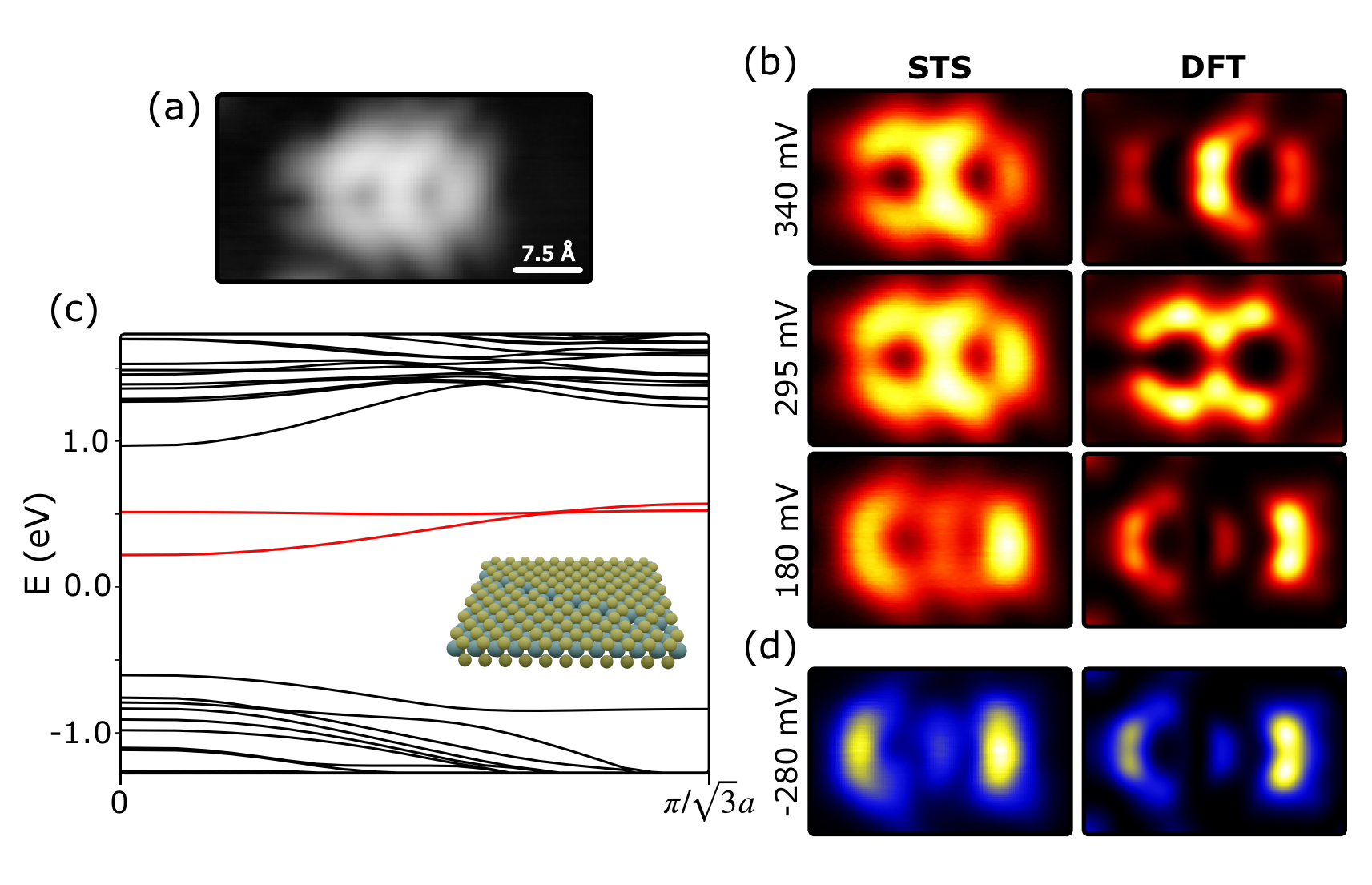}
	\caption{Artificial sulfur vacancy structures in \mo{}. (a) STM image of a sulfur vacancy dimer ($V_{\text{bias}}=$ 1 V, $I_{\text{set}}=$ 100 pA, image size: 4$\times$2 nm$^{2}$). (b) Constant-height d$I$/d$V$ maps measured at the indicated $V_{\rm bias}$ compared to DFT simulated DOS maps of the corresponding hybrid orbitals (image sizes: 2.2$\times$1.4 nm$^{2}$). (c) Band structure of a vacancy chain (model in inset). (d) Constant-height current map of the vacancy dimer in the charged state ($V_{\text{bias}}=-280$ mV).}
	\label{figure 4}
\end{figure}

Until now theoretical investigations only predicted the existence of one type of JT effect for negatively charged sulfur vacancies in \mo{}~\cite{tan2020stability,komsa2015native,gupta2018two}. Similarly, two recent experimental works on charged sulfur vacancies in single-layer \mo{} on Gr proposed only one type of distortion to explain the observed orbital symmetry reduction \cite{xiang2023charge, aliyar2024symmetry}. Our investigations, however, reveal that there are two distinct types of JT distortions, denoted JT1 and JT2. They correspond to two imaginary phonon modes and thus two types of lattice distortions. Both are local minima in the potential energy surface (PES) of the system, which is shown in Fig. \ref{figure 3}(c). A direct comparison between our experimental maps and DFT simulations of the HODO, shown in Fig. \ref{figure 2}(e), makes plain that the two orbitals denoted by ($\times$) and ($\ocircle$) correspond to JT1 and JT2. We thus indicate the local minima of the PES that correspond to JT1 and JT2 by crosses ($\times$) and circles ($\ocircle$), respectively. A cut through the PES along the horizontal line indicated in black, is shown in Fig. \ref{figure 3}(d). The JT distortion leads to an energy gain of approximately 50 meV compared to the unrelaxed structure, while JT1 and JT2 distortions have the same energy within the accuracy of our DFT calculation. Upon charging, the sulfur vacancy thus spontaneously transitions into one of the JT distorted states with similar probabilities (blue arrows). The corresponding structural distortions compared to the neutral vacancy configuration [Fig. \ref{figure 3}(e)] are shown in Fig. \ref{figure 3}(f) and (g). A detailed summary of all simulated JT1 and JT2 orbitals is presented in Fig. S5 and S6. 

Coming back to the transformation from JT1 to JT2 shown in Fig. \ref{figure 2}(f), we can now infer that it is likely caused by the strain field created when a neighboring vacancy becomes JT distorted. Indeed, DFT is able to reproduce the transformation using a directional strain field, see Fig. S7. Apparently the elastic energy is minimized, when the two neighboring vacancies are in different JT states. 

To demonstrate the potential of our vacancy creation technique for the study of larger tailor-made sulfur vacancy structures, we created and characterized a vacancy dimer, see Fig. \ref{figure 4}(a). The vacancies are created in next-nearest neighbor distance $d=\sqrt{3}a$, with $a$ being the \mo{} lattice constant. As can be seen in Fig. \ref{figure 4}(b) (left column) the d$I$/d$V$ intensity distribution of the hybrid orbitals are distinct from the LUDO and LUDO+1 of isolated vacancies. The observed behavior is reminiscent of bonding and anti-bonding molecular orbitals, which is reproduced in the corresponding DFT maps (right column). While the measured d$I$/d$V$ signal seems to suggest that the anti-bonding state is lower in energy, as also observed for coupled quantum corrals confining hole-like quasiparticles~\cite{jolie2022creating}, an inspection of the total DOS does not allow for such an unambiguous identification, see Fig. S8. %we stress here that such an identification is not unambiguous.} 
Nevertheless, these hybridized orbitals form two unoccupied bands in a one-dimensional vacancy chain, according to our DFT calculations presented in Fig. \ref{figure 4}(c). Such an assembly of large vacancy defect structures can be realized with our atomically precise vacancy creation method, if the adatoms used as markers can be manipulated laterally with sufficiently high fidelity prior to pick-up. The chains would have great potential for 2D devices made from semiconducting TMDs, in which transport properties are often governed by defect states \cite{qiu2013hopping, gali2020electronic}. While for Fe adatoms on \mo{} the manipulation efficiency is limited, it is plausible that several adatom species exist, which can be laterally manipulated, while simultaneously forming sufficiently strong bonds with the chalcogen atoms to enable pick-up.

Applying a negative bias voltage enables us to charge the dimer. Its HODO is shown in Fig. \ref{figure 4}(d), in good agreement with the corresponding DFT simulation for single electron charging. DFT simulations for all orbitals of the singly and doubly charged vacancy dimer are provided in Fig. S9. The LUDO of the neutral dimer (measured at 180 mV) and the HODO of the charged dimer (measured at -280 mV) have the same shape and hence display no JT distortion, in stark contrast to the spontaneous symmetry breaking observed in the charged single vacancy. This can be rationalized by the fact that the dimer geometry reduces the symmetry and consequently lifts the orbital degeneracy. It emphasizes the importance of the environment of single vacancies, to enable detection of their pristine properties.

In summary, we have developed a technique for atomically precise creation of point defects in 2D TMDs, involving adsorbed atoms and the tip of an STM. Sulfur vacancies in the top sulfur layer of single-layer \mo{} are created by picking up single Fe adatoms from the surface. Tip-induced gating was used to manipulate sulfur vacancies into a negatively charged state, revealing the existence of two types of JT distortions, which can be transformed into one another by their mutual interaction. Furthermore, we studied the orbital overlap of sulfur vacancy states using the described creation approach and revealed strong hybridization.

We acknowledge funding from Deutsche Forschungsgemeinschaft (DFG) through CRC 1238 (project number 277146847, subprojects A01 and B06). W. J. acknowledges financial support from the DFG through project JO 1972/2-1 within the SPP 2244. J. F. acknowledges financial support from the DFG SPP 2137 (project FI 2624/1-1). H.-P. K. thanks CSC-IT Center for Science Ltd. for generous grants of computer time. A. V. K. acknowledges funding from DFG through project KR 4866/9-1 and the collaborative research center "Chemistry of Synthetic 2D Materials" CRC-1415-417590517.

\appendix

\bibliographystyle{prstynoetal}
\bibliography{library.bib}

\end{document}

\DeclareGraphicsExtensions{.pdf}

\title{Supplemental Material:\\
Tip-induced creation and Jahn-Teller distortions of\\
sulfur vacancies in single-layer \mo{}
}

\author{Daniel Jansen}
\affiliation{II. Physikalisches Institut, Universit\"{a}t zu K\"{o}ln, Z\"{u}lpicher Stra\ss e 77, 50937 K\"{o}ln, Germany}
\author{Tfyeche Tounsi}
\affiliation{II. Physikalisches Institut, Universit\"{a}t zu K\"{o}ln, Z\"{u}lpicher Stra\ss e 77, 50937 K\"{o}ln, Germany}
\author{Jeison Fischer}
\affiliation{II. Physikalisches Institut, Universit\"{a}t zu K\"{o}ln, Z\"{u}lpicher Stra\ss e 77, 50937 K\"{o}ln, Germany}
\author{Arkady V. Krasheninnikov}
\affiliation{Institute of Ion Beam Physics and Materials Research, Helmholtz-Zentrum Dresden-Rossendorf, 01328 Dresden, Germany}
\author{Thomas Michely}
\affiliation{II. Physikalisches Institut, Universit\"{a}t zu K\"{o}ln, Z\"{u}lpicher Stra\ss e 77, 50937 K\"{o}ln, Germany}
\author{Hannu-Pekka Komsa}
\affiliation{Faculty of Information Technology and Electrical Engineering, University of Oulu, FI-90014 Oulu, Finland}
\author{Wouter Jolie}
\affiliation{II. Physikalisches Institut, Universit\"{a}t zu K\"{o}ln, Z\"{u}lpicher Stra\ss e 77, 50937 K\"{o}ln, Germany}

\maketitle

\newpage

\tableofcontents

\newpage

\section{Sulfur vacancy appearance in STM}

\begin{figure*}[htb]
	\centering
		\includegraphics[width=0.9\textwidth]{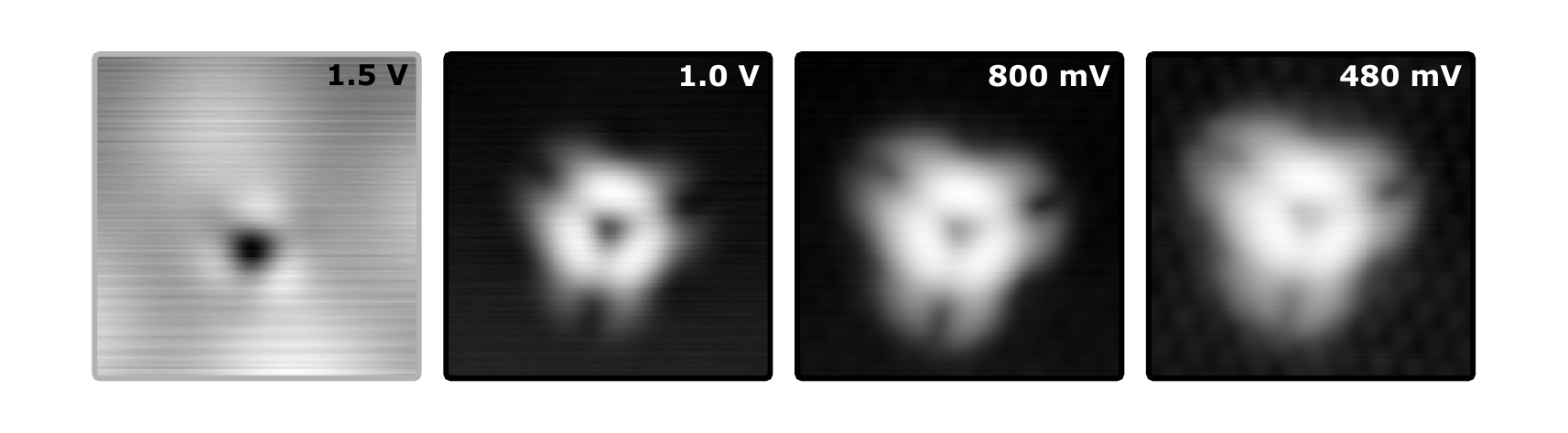}
	\caption{Bias dependence of sulfur vacancy appearance in STM topographies (I$_{\text{set}}$= 100 pA, image sizes: 3$\times$3 nm$^{2}$). At high bias voltages the sulfur vacancy appears as a local depression, at low bias voltages the three-fold symmetric vacancy lobes dominate the appearance.}
	\label{bias dependence}
\end{figure*}

The STM is sensitive to both surface corrugation and electronic structure. From a purely structural perspective, sulfur vacancies are expected to appear as local depressions in the STM images. From a purely electronic perspective however, one would expect to see the three-fold symmetric vacancy orbitals. Which contribution dominates the STM image strongly depends on the bias voltage applied. To illustrate this, a series of STM images of a sulfur vacancy measured at different bias voltages is presented in Fig. \ref{bias dependence}. 

\newpage

\section{On the use of constant-height current maps to access the HODO orbital}

In this section we explain in more detail our choice to measure the HODO using constant-height current maps instead of using d$I$/d$V$ maps. The latter give us access to the DOS at the applied bias voltage, which can be directly converted to an energy $E=e V_{\rm bias}$. Constant-height current maps, however, integrate over all available states between the bias voltage and the Fermi level (at $V_{\rm bias}=0$).

\begin{figure*}[htb]
	\centering
		\includegraphics[width=0.9\textwidth]{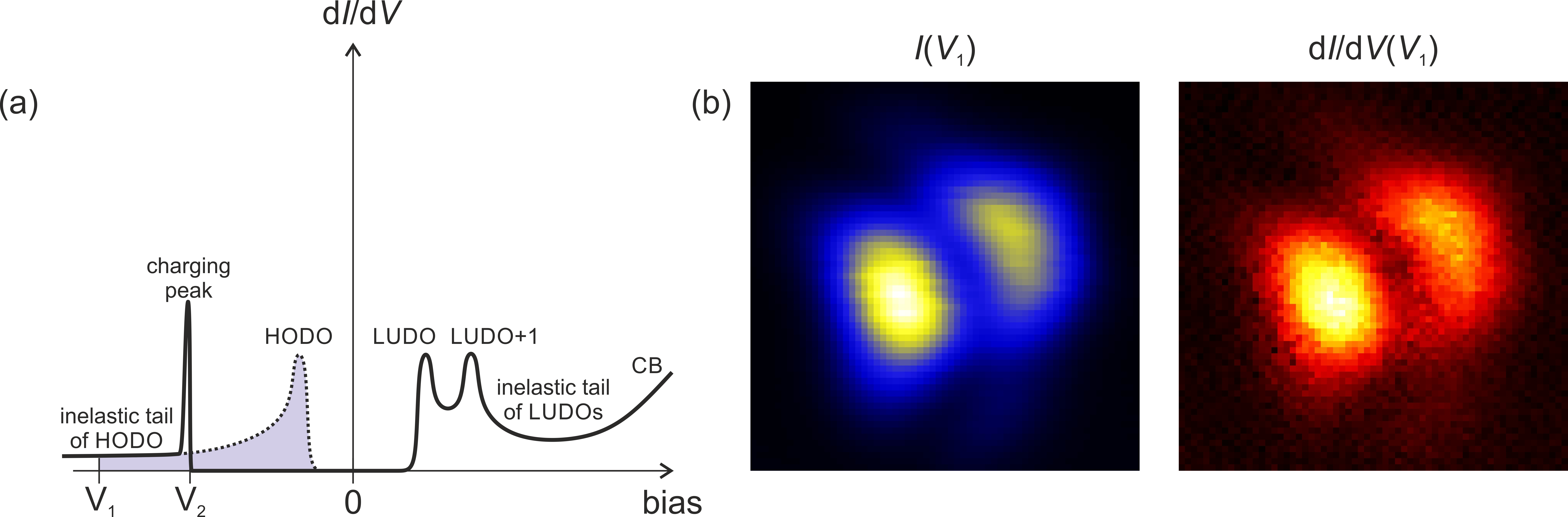}
	\caption{Reasoning behind constant-height current maps. (a) Sketch illustrating a typical d$I$/d$V$ spectrum measured on a sulfur vacancy in \mo{} (solid line). Biases $V_1$ and $V_2$ correspond to significant features in the spectrum, see text. The dotted line shows the HODO state of a negatively charged vacancy, which emerges when the bias drops below $V_2$, but cannot be measured by d$I$/d$V$ at its peak energy. (b) Constant-height current and d$I$/d$V$ maps (2$\times$2 nm$^{2}$) measured at $V_1=-1$~V.
 }
	\label{current maps sketch}
\end{figure*}

Fig. \ref{current maps sketch}(a) sketches as a black solid line the main features in d$I$/d$V$ spectra measured on a vacancy. At positive bias, we find the LUDO and LUDO+1 orbitals, followed by an inelastic tail often observed for in-gap states in semiconducting TMDs~\cite{schuler2019large, cochrane2021spin}. The conduction band (CB) of \mo~leads to an additional increase at high positive bias voltages. At negative bias voltages, we find the charging peak ($V_2$), followed by a finite d$I$/d$V$ signal ($V_1$). This signal stems from the inelastic tail of the HODO, as explained below. The charging bias $V_2$ strongly depends on the position we acquire the spectrum. It leads to a pronounced ring in d$I$/d$V$ maps separating the areas in which the defect is charged or neutral.

The charging event represents the bias at which the LUDO is filled by an electron, thereby becoming the HODO. The HODO is, however, still between the Fermi level and the applied bias voltage ($V_{\rm{bias}}< -500$ mV). In consequence, the HODO will not appear in the d$I$/d$V$ spectrum, since its energy is below the charging energy. Only its inelastic tail will appear at bias voltages below the charging peak, as sketched close to $V_1$ in Fig. \ref{current maps sketch}(a). Because we use the tip as a tool to charge the vacancy, we are not more free to conduct d$I$/d$V$ spectroscopy at the voltage at which the HODO is located.

Since d$I$/d$V$ can only access the (small) inelastic tail of the HODO, one might want to integrate over the entire spectral intensity instead of only measuring the tail of the peak under investigation. Indeed, the peak of the HODO is accessible, when mapping the tunneling current in constant-height mode. When applying a high negative bias ($V_1$) we are able to map the HODO in the charged state, see Fig. \ref{current maps sketch}(b). This map represents the integrated intensity at negative bias, shown as shaded area in Fig. \ref{current maps sketch}(a). As a direct comparison, we also present the d$I$/d$V$ map measured simultaneously in Fig.~\ref{current maps sketch}(b). Though the signal-to-noise ratio is lower, it has strong resembles with the current map, in line with our interpretation that it captures the tail of the HODO. Due to better signal-to noise ratio it is favorable to choose constant-height current maps instead of d$I$/d$V$ maps. An additional advantage is  that the charging peak does not influence the measurement, when the bias is sufficiently negative to charge the defect in the entire frame of the map.

\newpage

\section{Energetic shift of SOC-split states}

\begin{figure*}[htb]
	\centering
		\includegraphics[width=0.9\textwidth]{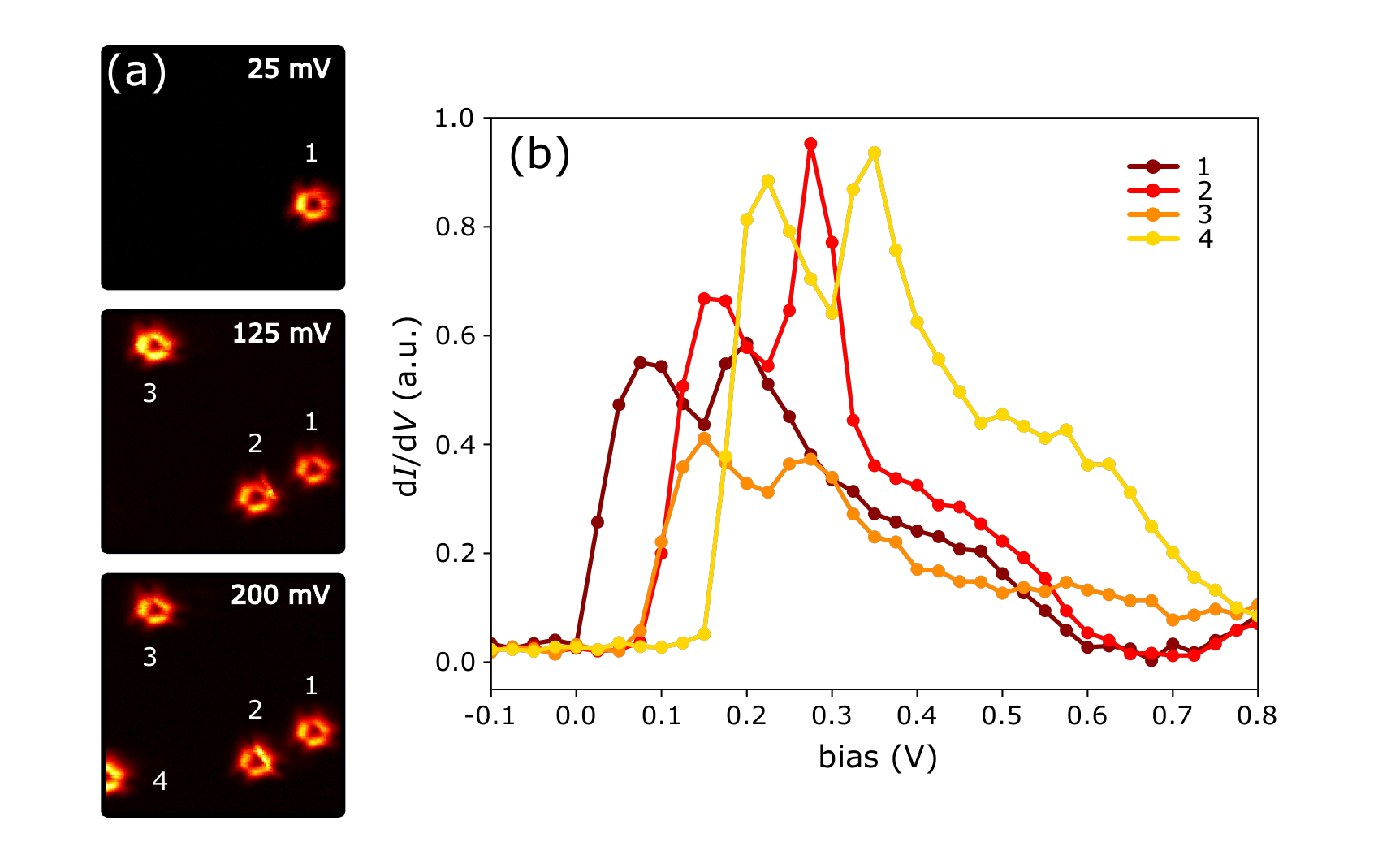}
	\caption{Variation of LUDO and LUDO+1 energies of different neutral sulfur vacancies. (a) d$I$/d$V$ maps of the same area containing four sulfur vacancies measured at different bias voltages. The characteristic three-fold symmetric sulfur vacancy orbitals appear successively with increasing bias voltage depending on the positioning of the respective vacancy states. (b) d$I$/d$V$ point spectra ($V_{\text{stab}}=$ 1.3 V, $I_{\text{stab}}=$ 200 pA, $V_{\text{mod}}=$ 1 mV) measured on the lobes of the four sulfur vacancies shown in (a). The positions of the two peaks in the spectra reveal shifts of the LUDO and LUDO+1 states in energy for different vacancies.}
	\label{SOC-state shifts}
\end{figure*}

The position of the SOC-split states of different neutral sulfur vacancies is found to vary in energy. This observation can be explained by band bending at islands edges and (line) defects \cite{murray2020band}. Measurements are presented in Fig. \ref{SOC-state shifts}. In (a), d$I$/d$V$ maps of an area on a \mo{} island with four sulfur vacancies is shown. In the maps, the vacancies appear at different energies, which is due to the different energetic positions of their SOC-split states in the \mo{} band gap. This becomes apparent in the point spectra acquired on the lobes of the vacancies and is presented in Fig. \ref{SOC-state shifts}(b).

\newpage

\section{Computational methods}

All density-functional theory calculations were carried out within the projector-augmented wave formalism as implemented in the VASP software package \cite{vasp1,vasp2,vasp3}. We used Perdew-Burke-Ernzerhof (PBE) functional with van der Waals interactions accounted via Grimme's D2 corrections and plane-wave basis defined by 500 eV cutoff \cite{pbe,grimmed2}. The defects were modeled with a 6$\times$6 supercell and the Brillouin zone sampled using a 2$\times$2 k-point mesh. The vibrational modes were evaluated using phonopy \cite{phonopy}.

\section{Simulated STM images and partial charge density plots of single vacancies}

Simulated STS maps and charge density plots were computed for neutral and charged sulfur vacancies using DFT. The maps were simulated with Tersoff-Hamann scheme from the partial DOS at about 3 {\AA} from the outermost plane of S atoms. These calculations always included spin-orbit coupling and the defect wave functions were extracted using vaspkit \cite{vaspkit}. Note that only $\Gamma$-point states are shown. The positions of atoms in the \mo{} lattice are overlaid and indicated by green crosses for Mo and cyan dots for sulfur atoms. 

The Fermi-level position in the DOS plots is arbitrarily set between the highest occupied and lowest unoccupied states in the system. In experiment, the Fermi-level position is additionally governed by the Gr/Ir(111) substrate and the tip potential, which are not included in our model.

The \mo~band gap is underestimated when using semilocal functionals, such as PBE adopted herein. In our preliminary tests with HSE functional, the occupied defect states of charged vacancies were pushed markedly lower in energy, but had a fairly small effect on the unoccupied state energies and negligible effect on the simulated STM images. Since we cannot directly compare the occupied state energy to experiments or probe the splitting of the unoccupied states of a charged vacancy, we chose to use PBE due to more manageable computational cost. It is also worth noting that (i) all considered defect states are completely inside the band gap and spatially well localized and (ii) SOC-induced splitting of defect states tends to be rather well described by DFT~\cite{schuler2019large}.

\begin{figure*}[htb!]
    \begin{center}
       \includegraphics[width=16cm]{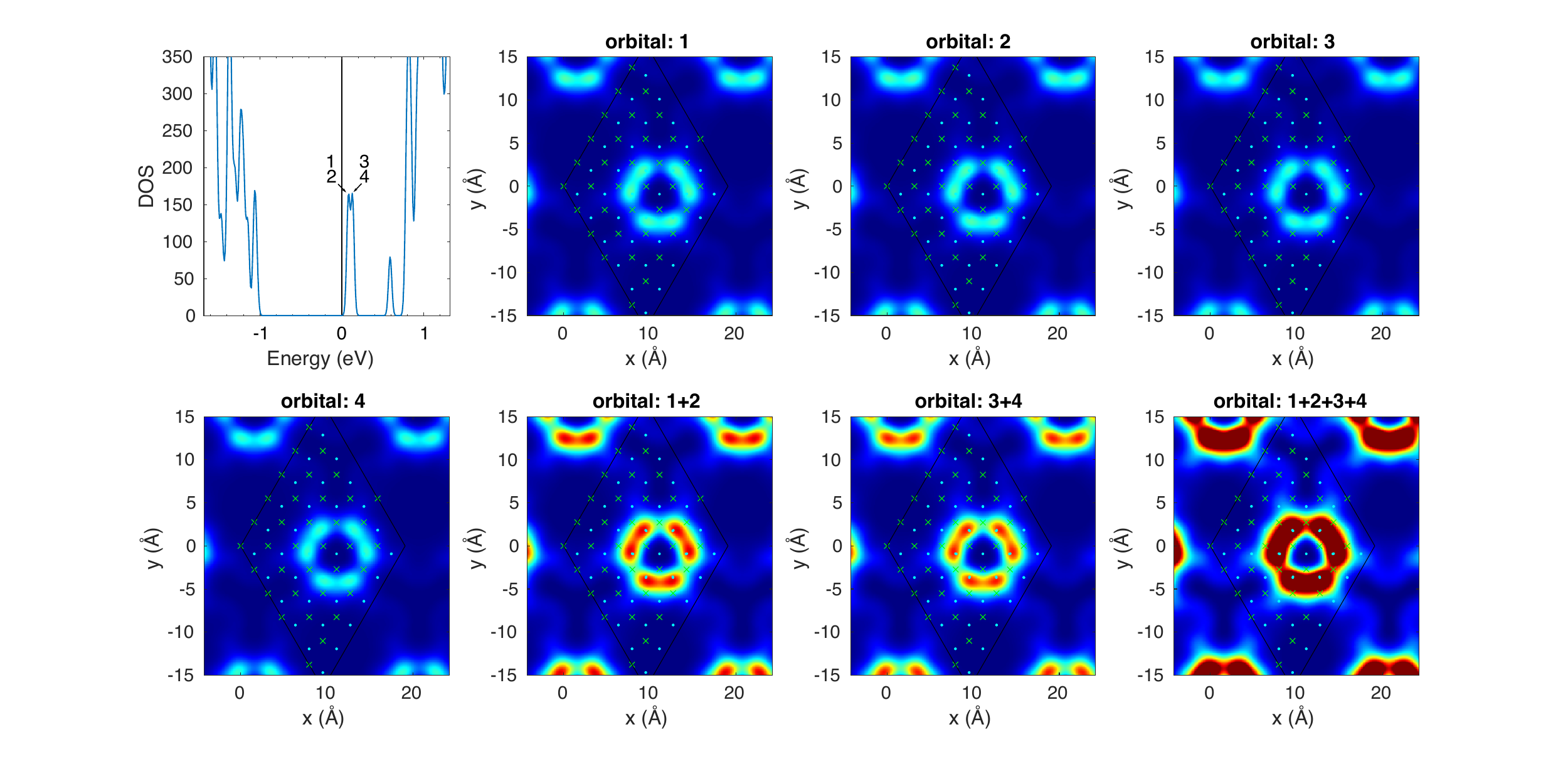}\\[\smallskipamount]
       \includegraphics[width=16cm]{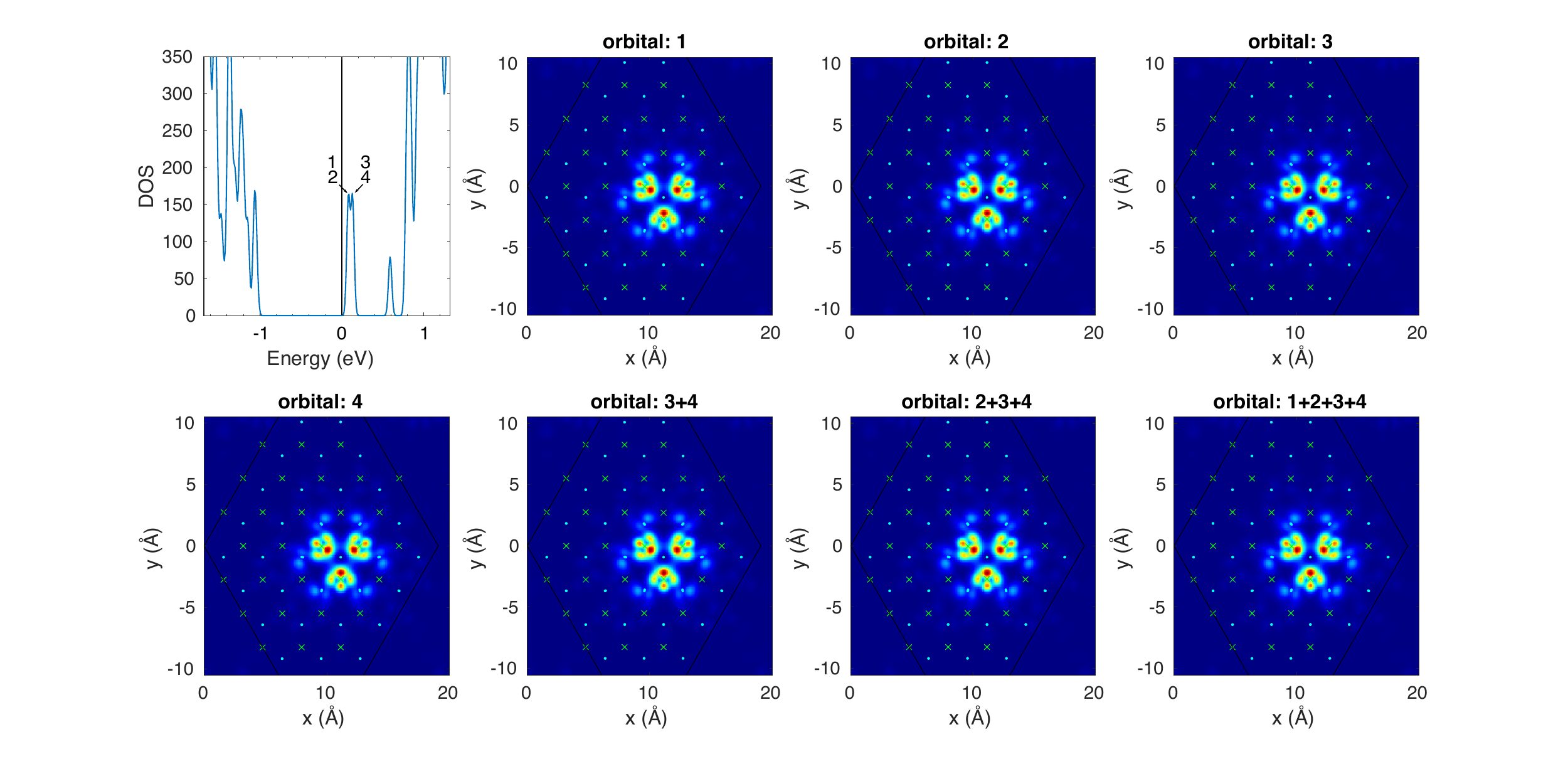}
     \end{center}
     \caption{
     Simulated STM images (first set) and total integrated charge densities (second set) of the neutral sulfur vacancy. The first panel shows the DOS, in which orbitals are indicated by numbers. The Fermi level is represented by the black vertical line. The atomic positions of the \mo{} lattice are overlaid (green crosses for Mo and cyan dots for sulfur atoms).}
     \label{single neutral vac}
\end{figure*}

\begin{figure*}[htb!]
\begin{center}
  \includegraphics[width=16cm]{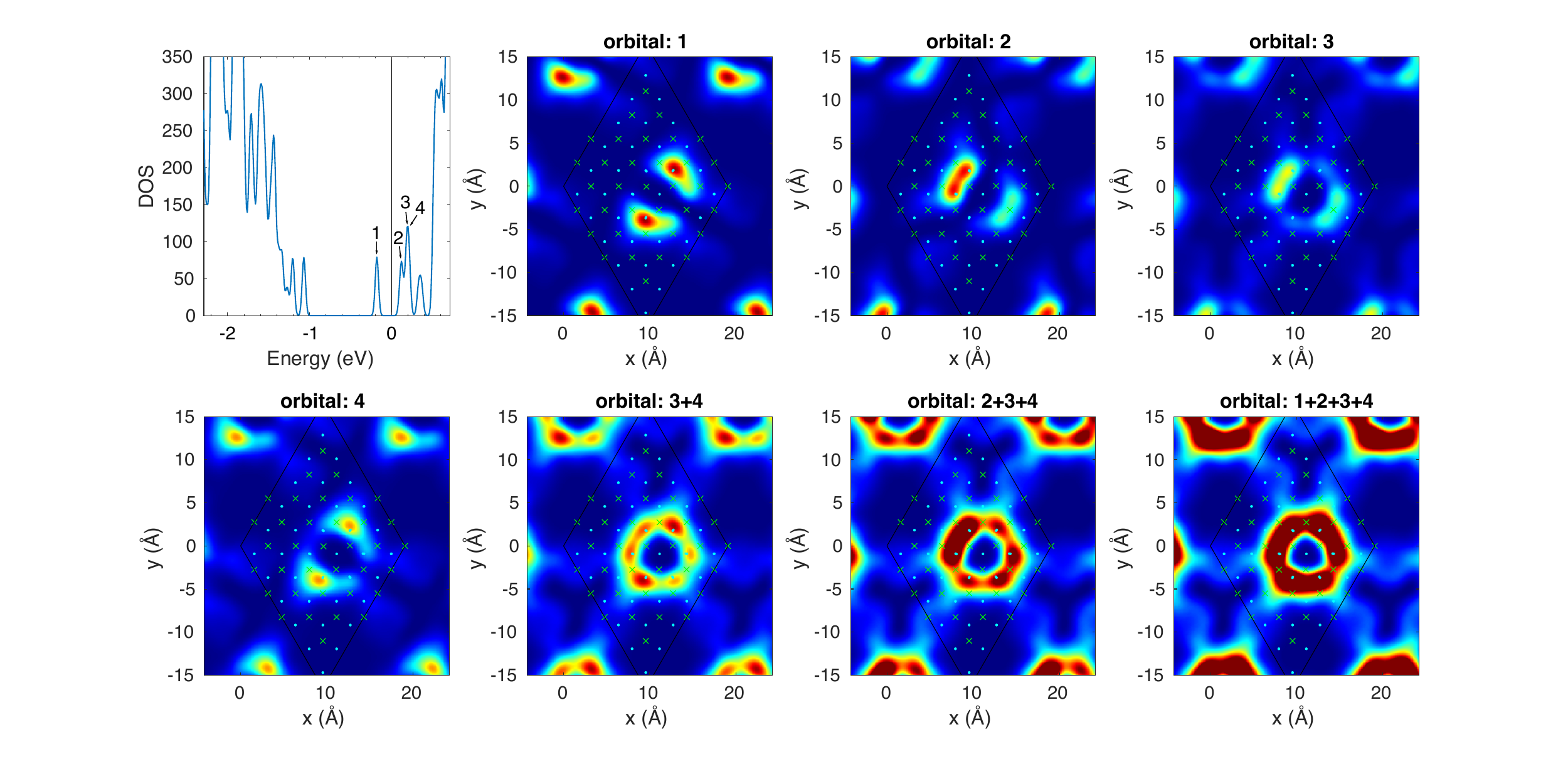}\\[\smallskipamount]
  \includegraphics[width=16cm]{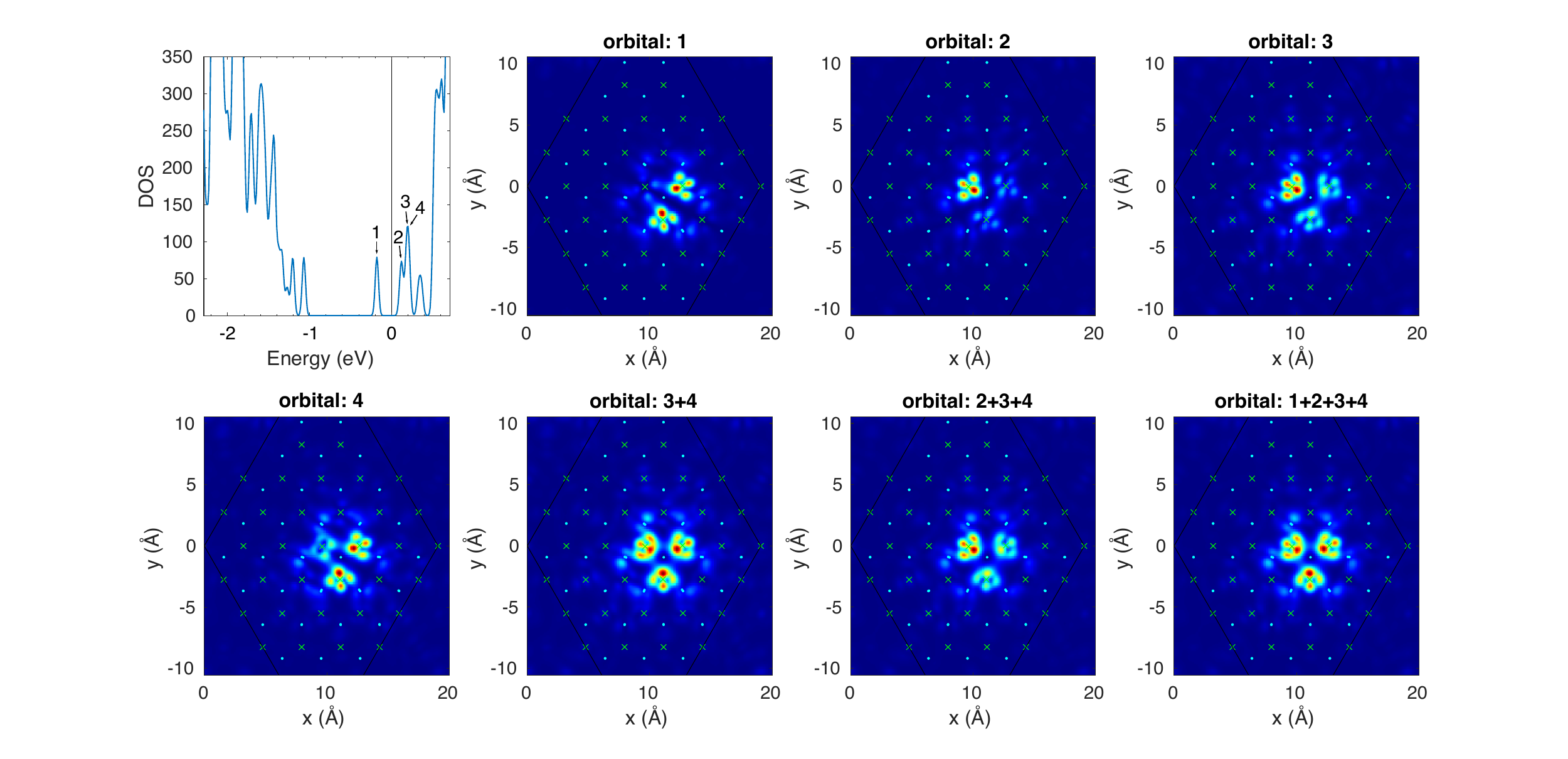}
\end{center}
\caption{
 Simulated STM images (first set) and total integrated charge densities (second set) of the negatively charged sulfur vacancy in the JT1 state. The first panel shows the DOS, in which orbitals are indicated by numbers. The Fermi level is represented by the black vertical line. The atomic positions of the \mo{} lattice are overlaid (green crosses for Mo and cyan dots for sulfur atoms). From the charge density plots, the orbitals can be confirmed to have the same form as obtained from symmetry-adapted linear combinations given in the main paper. Here orbital 1 corresponds to $\phi_3$ and and orbital 2 to $\phi_2$. Only orbital 1 is accessible in our experiments.}
 \label{single charged vac JT1}
\end{figure*}

\begin{figure*}[htb!]
\begin{center}
  \includegraphics[width=16cm]{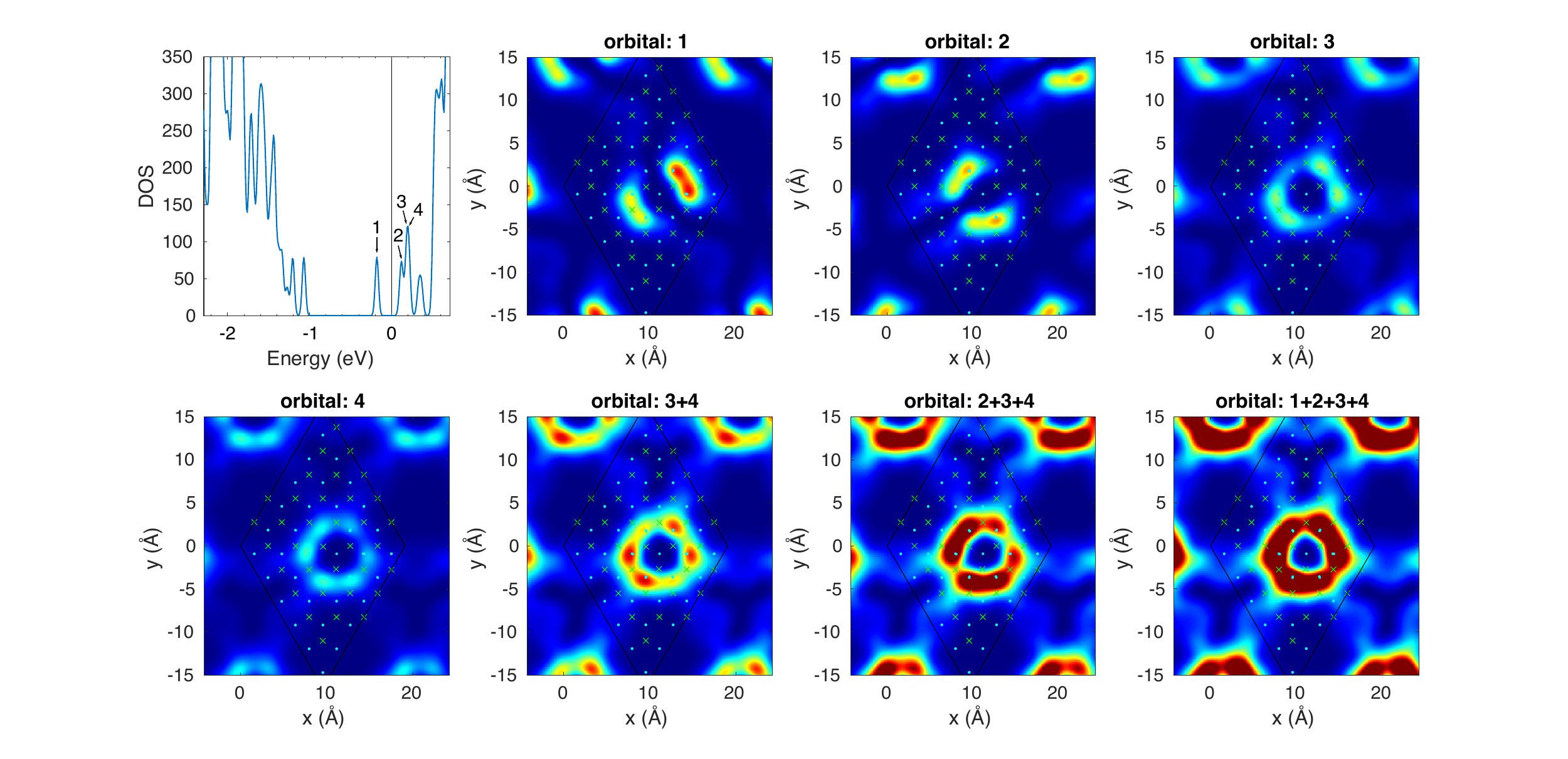}\\[\smallskipamount]
  \includegraphics[width=16cm]{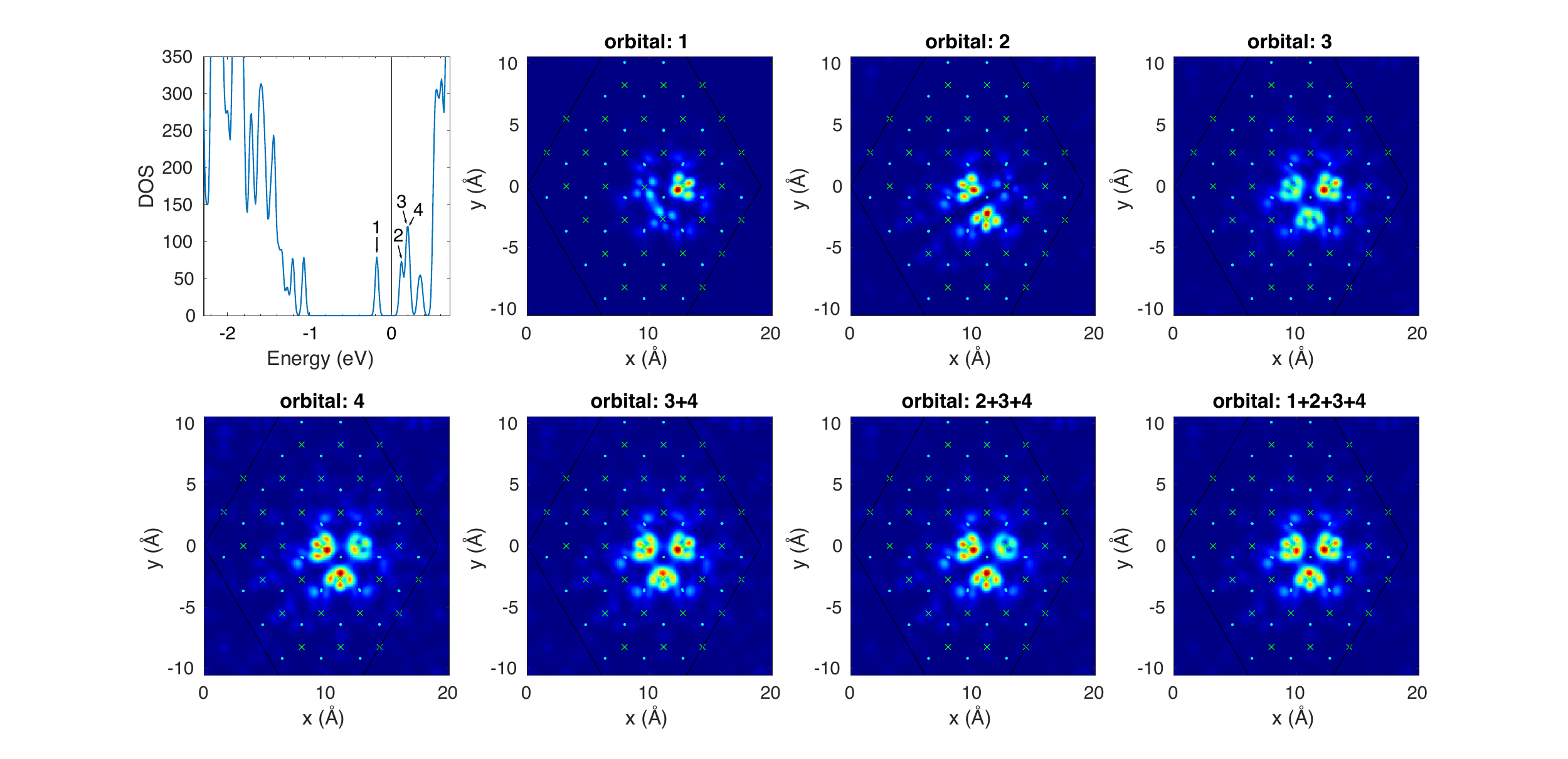}
\end{center}
\caption{
Simulated STM images (first set) and total integrated charge densities (second set) of the negatively charged sulfur vacancy in the JT2 state. The first panel shows the DOS, in which orbitals are indicated by numbers. The Fermi level is represented by the black vertical line. The atomic positions of the \mo{} lattice are overlaid (green crosses for Mo and cyan dots for sulfur atoms). Only orbital 1 is accessible in our experiments.}
\label{single charged vac JT2}
\end{figure*}

Data for a single neutral vacancy is presented in Fig. \ref{single neutral vac}. All four orbitals appear very similar in the simulations. Results for single negatively charged sulfur vacancies are shown in Fig. \ref{single charged vac JT1} for the case of JT1 and in Fig. \ref{single charged vac JT2} for JT2. Note that the states have appearances expected from SALCs, except that the order of the states is swapped in the two JT configurations.

\newpage

\newpage

\newpage

\section{Effect of lattice strain on Jahn-Teller distortions}

\begin{figure*}[htb]
	\centering
		\includegraphics[width=0.6\textwidth]{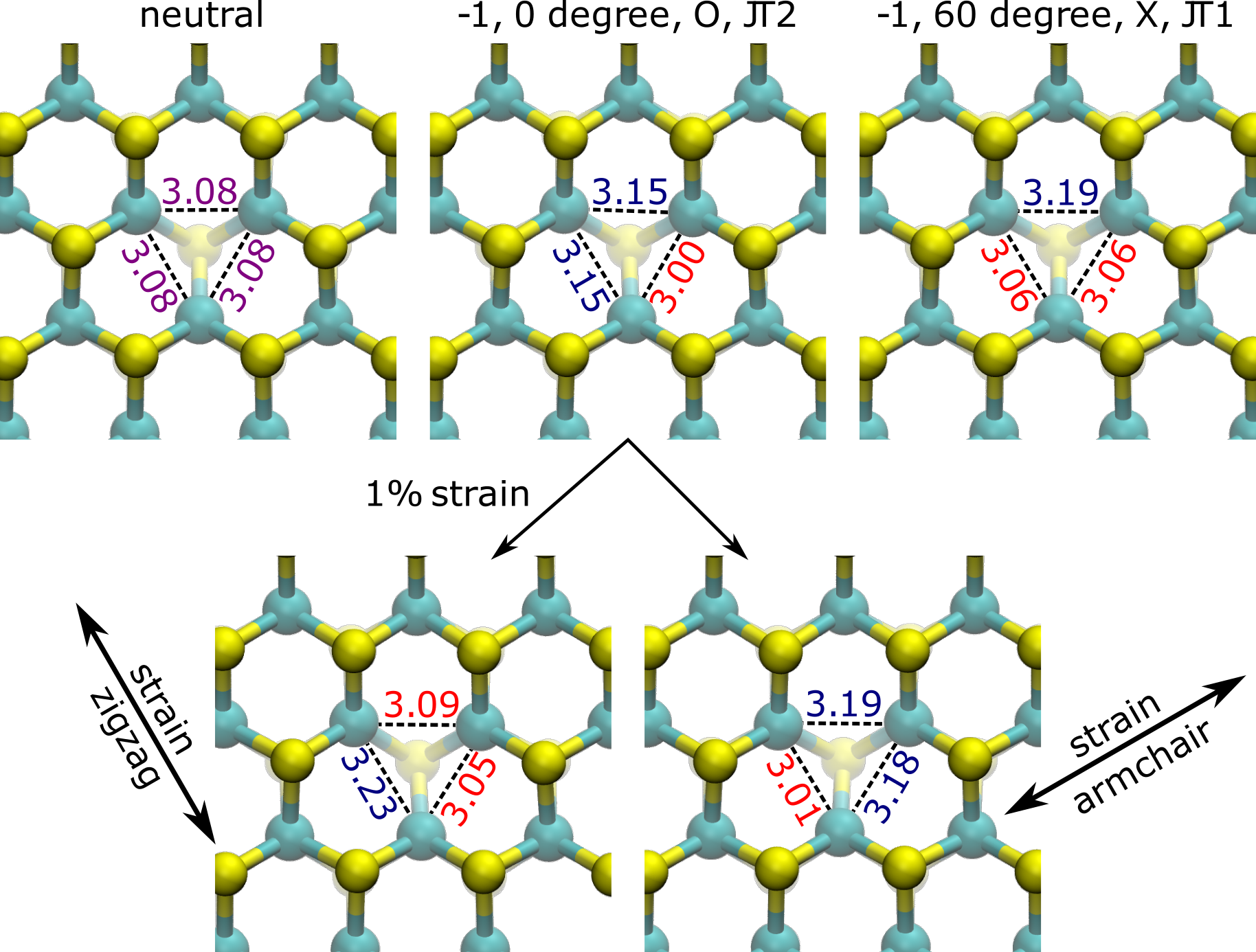}
	\caption{Effect of uniaxial strain on the Jahn-Teller effect in charged sulfur vacancies. Application of one percent of uniaxial strain along the armchair direction of the \mo~lattice leads to a change from the atomic configuration associated with JT2 distortion to the atomic configuration associated with the JT1 distortion. The numbers on the dashed lines around the sulfur vacancy give the Mo atom separation (\AA).}
	\label{figure strain}
\end{figure*}

To understand the influence of strain on the atomic configuration around charged sulfur vacancies and their associated Jahn-Teller distortions, we performed DFT calculations for application of uniaxial strain along the armchair and zigzag directions of the \mo~lattice. Results for the application of 1 \% of strain on the JT2 distorted lattice is shown in Fig. \ref{figure strain}. The numbers on the dashed lines around the sulfur vacancy give the Mo atom separation in units of \AA. Red numbers indicate smaller distances and blue numbers larger distances than in the configuration for a neutral sulfur vacancy. It becomes apparent that  1 \% strain along the zigzag direction is sufficient to transform the atomic configuration associated with the JT2 distortion into the configuration associated with the JT1 distortion. The presented theoretical data therefore evidences the possibility of transitioning from one JT distorted state to the other by application of strain, e.g. by charging of a nearby sulfur vacancy. It therefore supports the experimental findings presented in Fig. 2(f) in the main manuscript.

\newpage

\section{Simulated STM images and partial charge density plots of vacancy dimer}

\begin{figure*}[htb!]
\begin{center}
  \includegraphics[width=16cm]{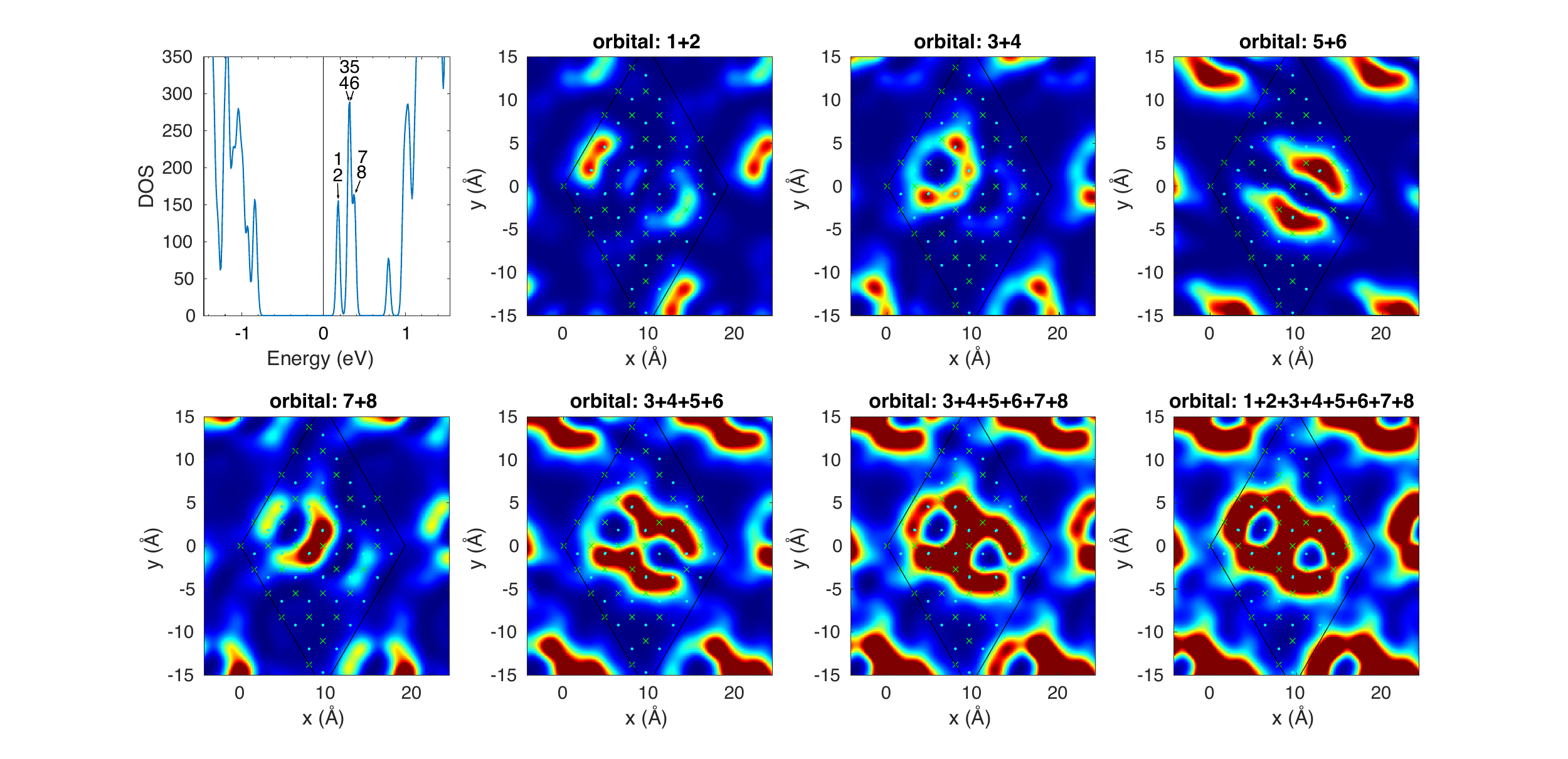}\\[\smallskipamount]
  \includegraphics[width=16cm]{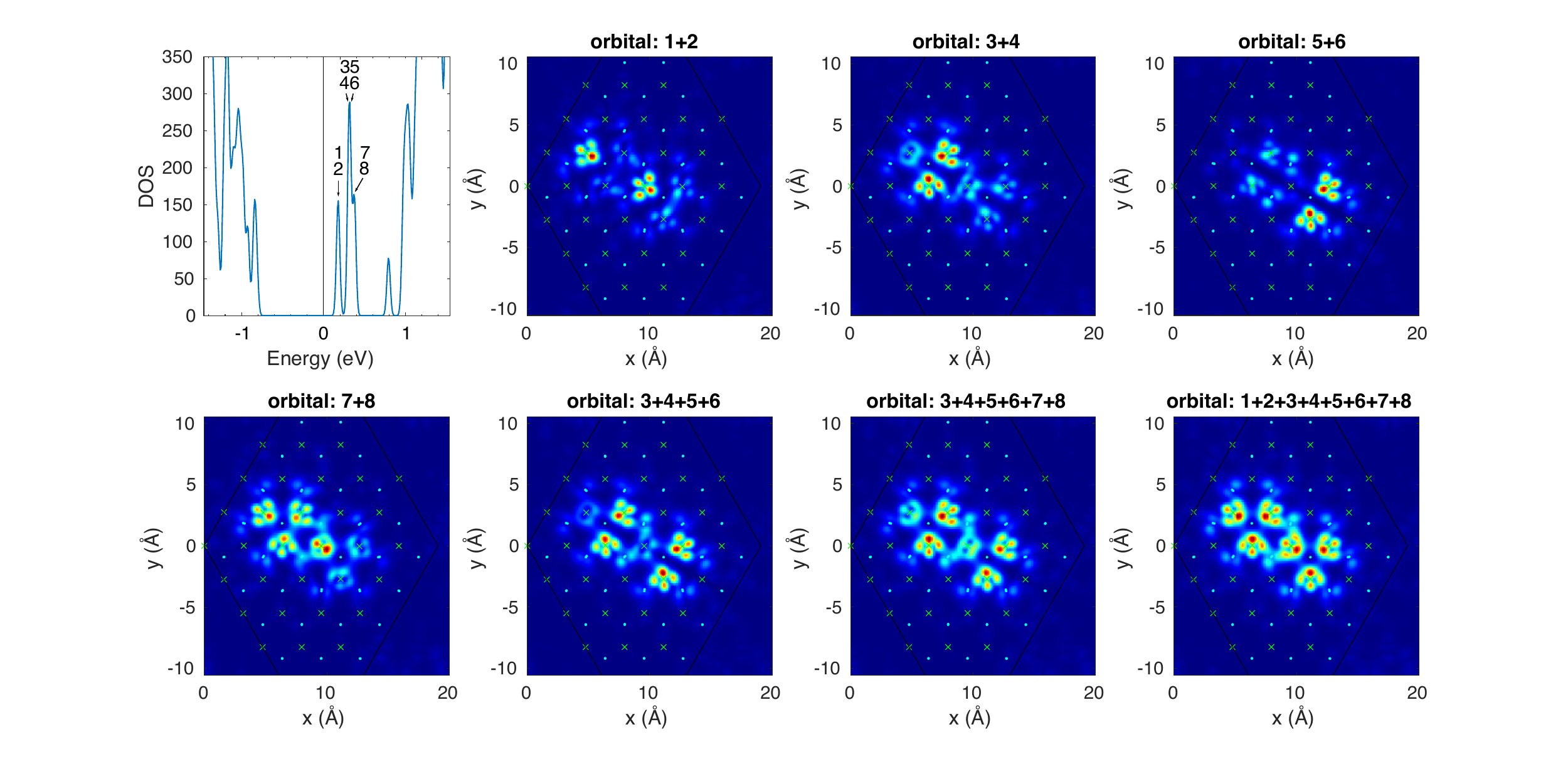}
\end{center}
\caption{
Simulated STM images (first set) and total integrated charge density (second set) of a sulfur vacancy dimer in the $\sqrt{3}a$-configuration in the neutral charge state. The first panel shows the DOS, in which orbitals are indicated by numbers. The Fermi level is represented by the black vertical line. The atomic positions of the \mo{} lattice are overlaid (green crosses for Mo and cyan dots for sulfur atoms).}
\label{vac dimer neutral}
\end{figure*}

\begin{figure*}[htb!]
\begin{center}
  \includegraphics[width=16cm]{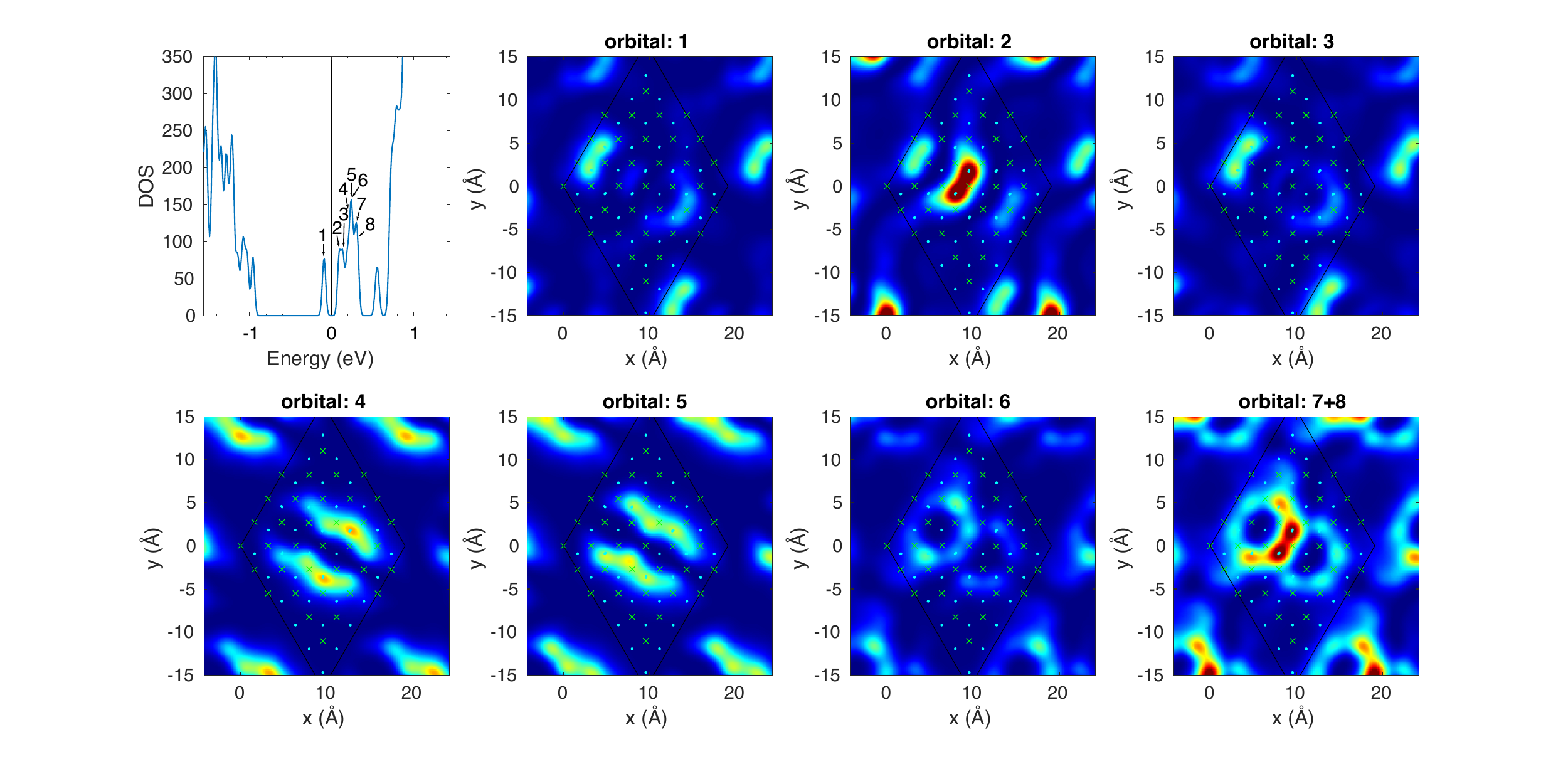}\\[\smallskipamount]
  \includegraphics[width=16cm]{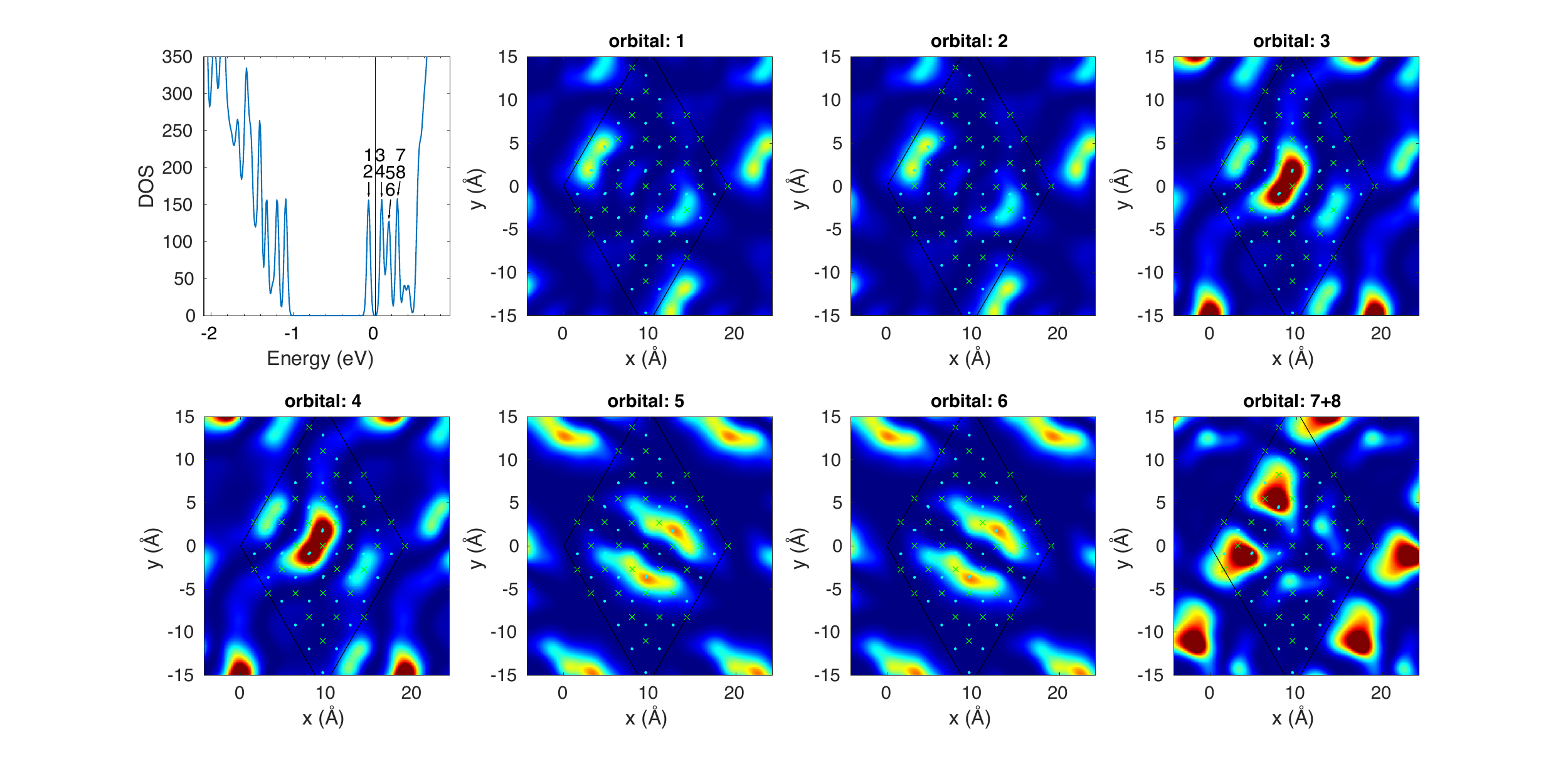}
\end{center}
\caption{
Simulated STM images of a sulfur vacancy dimer in the $\sqrt{3}a$-configuration in singly (first set) and doubly (second set) charged state. The first panel shows the DOS, in which orbitals are indicated by numbers. The Fermi level is represented by the black vertical line. The atomic positions of the \mo{} lattice are overlaid (green crosses for Mo and cyan dots for sulfur atoms).}
\label{vac dimer charged}
\end{figure*}

\clearpage
\appendix
\bibliographystyle{prstynoetal}
\bibliography{./sm_library.bib}